Computer Systems
Science and Engineering

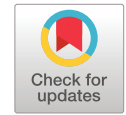

Tech Science Press

# Optimizing Region of Interest Selection for Effective Embedding in Video Steganography Based on Genetic Algorithms

## Nizheen A. Ali[1] and Ramadhan J. Mstafa[2,3,*]

[1]Department of Computer Science, College of Science, University of Duhok, Duhok, 42001, Iraq
[2]Department of Computer Science, Faculty of Science, University of Zakho, Zakho, 42002, Iraq
[3]Department of Computer Science, College of Science, Nawroz University, Duhok, 42001, Iraq
*Corresponding Author: Ramadhan J. Mstafa. Email: ramadhan.mstafa@uoz.edu.krd

**Abstract:** With the widespread use of the internet, there is an increasing need to ensure the security and privacy of transmitted data. This has led to an intensified focus on the study of video steganography, which is a technique that hides data within a video cover to avoid detection. The effectiveness of any steganography method depends on its ability to embed data without altering the original video's quality while maintaining high efficiency. This paper proposes a new method to video steganography, which involves utilizing a Genetic Algorithm (GA) for identifying the Region of Interest (ROI) in the cover video. The ROI is the area in the video that is the most suitable for data embedding. The secret data is encrypted using the Advanced Encryption Standard (AES), which is a widely accepted encryption standard, before being embedded into the cover video, utilizing up to 10% of the cover video. This process ensures the security and confidentiality of the embedded data. The performance metrics for assessing the proposed method are the Peak Signal-to-Noise Ratio (PSNR) and the encoding and decoding time. The results show that the proposed method has a high embedding capacity and efficiency, with a PSNR ranging between 64 and 75 dBs, which indicates that the embedded data is almost indistinguishable from the original video. Additionally, the method can encode and decode data quickly, making it efficient for real-time applications.

## 1  Introduction

As technology continues to advance at a rapid pace, individuals and organizations are becoming increasingly reliant on the Internet and various forms of digital communication. In light of this trend, the security of data transmission has become a crucial concern, particularly in contexts where confidential information is being shared. To address this issue, various techniques have been developed and implemented, including data hiding and cryptography. These methods are intended to protect digital data from unauthorized access and ensure the integrity and confidentiality of communications [1,2]. Cryptography and steganography are both techniques used to secure data transmission. Cryptography involves the use of mathematical algorithms to encrypt data, making it unreadable to unauthorized parties. The encryption process uses a key, which is known only to the sender and the intended recipient, to encode the data. On the other hand, steganography is the practice of hiding data within media files, such as images or audio files, in a way that is not immediately apparent to the observer. The data is embedded in the file in such a way that it is not easily detected or extracted, making it useful for secure communication [3]. Both cryptography and steganography can be used together to provide an additional layer of security for data transmission [4].

Video, as a medium, presents unique challenges in terms of data hiding and security due to its visual nature and the rapid advancement of technology in terms of video sharing and transmission. The human visual system's inability to detect small changes in a video makes it a particularly attractive medium for covert data transmission. The proliferation of video-sharing platforms and transmission methods exacerbates these security concerns [5]. In the field of steganography, it is crucial to consider the following three key factors: robustness, imperceptibility, and embedding capacity. Robustness pertains to the resilience of the steganography technique against signal processing and hacking attempts. Imperceptibility, on the other hand, relates to the degree of efficiency in which data is embedded, as a higher efficiency implies a lower likelihood of detection by potential hackers. Lastly, embedding capacity refers to the amount of data that can be concealed within a cover video [6].

Video steganography, specifically, can be categorized into two main types: spatial domain steganography and transform domain steganography. Spatial domain techniques involve direct manipulation of the pixel values in the cover video frames to hide data. Examples of spatial domain techniques include bit insertion and noise manipulation. On the other hand, transform domain techniques take advantage of the specific characteristics of video frames to embed data. These techniques involve transforming the frame from the spatial domain to the frequency domain using various transforms such as Discrete Cosine Transform (DCT), Discrete Fourier Transform (DFT), Discrete Wavelet Transform (DWT), Curvelet Transform, and Contourlet Transform [7]. The data is then hidden in the coefficients of the transformed video frame instead of the direct pixel values. Finally, the transformed frame is retransformed back to the spatial domain. Transform domain techniques offer several advantages over spatial domain techniques, including better imperceptibility, higher capacity, and increased security. However, these techniques are computationally more complex than spatial domain techniques and require more processing power. Therefore, the choice of technique depends on the specific requirements of the application [8].

In the field of video steganography, a further classification can be made based on the compression domains used in the process. Specifically, video steganography techniques can be categorized into two distinct categories: lossless and lossy. Lossless video steganography preserves both the original video and the concealed message in its original form, while lossy video steganography allows for some degradation of the original video to conceal the message. The latter method is particularly useful in situations where the original video is of lower quality or resolution, as the degradation caused by the steganography is less noticeable [9].

This paper presents a novel approach aimed at devising an optimized methodology for concealing confidential information within a carrier video frames with the objectives of enhancing both the payload capacity and quality. The proposed method integrates the AES encryption scheme, carrier video encoding, and the GA for data embedding. AES encryption provides a high level of security for the payload data, while carrier video encoding serves as a cover for embedding the encrypted data. The GA optimization algorithm optimizes the embedding process by selecting the optimal locations for embedding the encrypted data. This integration of techniques is expected to improve the effectiveness and efficiency of multimedia data hiding systems while also offering enhanced protection against potential attacks.

The following are the contributions of the proposed video steganography method:

- Incorporation of GA: The proposed method employs GA to improve the embedding capacity and security of the stego video.
- Optimization of the embedding process: The proposed method optimizes the embedding process by considering several important parameters such as the PSNR, MSE, embedding rate, and complexity of the embedding process.
- Enhanced Security: The proposed method enhances security by using GA to choose the optimal embedding locations and adjust the embedding parameters. This makes the stego video more resistant to attack and improves the security of the hidden data.
- Performance Evaluation: The proposed method includes a comprehensive performance evaluation that compares the effectiveness of the proposed video steganography method using a genetic algorithm and its variations without the use of the genetic algorithm.

The following sections of this paper are structured as follows: Section 2 provides an overview of the related works in video steganography and discusses the advantages and limitations of existing methods. Section 3 involves the use of AES encryption. Section 4 describes the proposed video steganography method, which involves carrier video encoding and decoding data in the stego video. Section 5 presents the experimental evaluation results of the proposed method. Finally, Section 6 concludes the paper.

## 2  Related Works

In recent literature, video steganography has garnered significant attention from researchers due to its potential for hiding large amounts of sensitive information in digital videos. One suggested approach, developed by Abbas et al. [10] utilizes the Cuckoo Search (CS) algorithm to optimize the process of video steganography. The method operates by extracting secret information byte byte and organizing the bits of each byte to form different variations. The CS algorithm is then used to search for the optimal pixel within the covering frame to hide the secret information. The Euclidean distance is used to compare the similarity between the pixels and the various byte forms to identify the optimal pixel. Additionally, the Levy fly random walk is employed to generate random movement from one pixel to another, and the 3-3-2 Least Significant Bit (LSB) replacement approach is used to identify the appropriate carrier pixel and incorporate the hidden information into its RGB components. The experimental results of this approach demonstrate high embedding efficiency, with a PSNR over 47 dB, as well as good visual quality as measured by PSNR and MSE measurements.

Also, Suttichaiya et al. [11] presented a technique for video steganography that uses AVI uncompressed video files to conceal text, images, audio, and video data. The method involves determining the size and number of frames and bytes to be used for concealment before loading the video file. The secret file is then transformed into binary using the dec2bin command and embedded into the video frames using the LSB method. The LSB format is chosen based on the color detection technique and the color's PSNR. However, this method is limited in its effectiveness when applied to grayscale or videos with an excessive amount of grayscale colors. Then, Suresh et al. [12] presented a new video steganography approach based on Oppositional Grey Wolf Optimization (OGWO). The method utilizes the discrete cosine transform (DCT) to identify scene changes and select the optimum frame for data concealment. The OGWO algorithm is then employed to discover the optimal locations for inserting private information within the selected frame. The method also employs Discrete Wavelet Transform (DWT) to create an ideal region for embedding the hidden data, and Inverse DWT to normalize the payload and video to improve the overall video quality. The proposed approach is a significant contribution to the field of steganography as it combines the use of OGWO and DWT to improve the efficiency and effectiveness of data concealment in video files.

In addition, Selim et al. [13] presented a system that combines both cryptography and steganography to enhance the security of concealed data. The confidential data is first encrypted using the AES encryption method before being embedded in a video file using steganography techniques. The authors chose video as the cover media due to its capacity to contain large amounts of data, and the FLV video format was selected for its relatively small file size compared to other formats. The system is capable of embedding both images and text using two methods: LSB and GA. The LSB method replaces the least significant bits of the cover frame with confidential data, while the GA employs a deep GA to choose the optimal locations for data concealment and determine the optimum frame based on the maximum fitness function possible and the smallest possible difference between the pixels of the cover frame and the secret data. The results of the study indicate that the cover video's size and the hidden image's quality remain unaffected by the data embedding, and the PSNR lies between 42 and 65 dBs. However, the study is limited in that it only demonstrates the outcome of data embedding in a single frame and does not examine the results of utilizing the entire video or explain the approach used to determine the optimum frame. Also, Abbasi-khazaei et al. [14] introduced an algorithm for virtual machine placement, which utilizes a multi-objective approach to minimize energy costs and optimize scheduling. According to the simulation results obtained on the CloudSim platform, this method has the capability to decrease energy costs, carbon footprints, service-level agreement violations, as well as the total response time for IoT requests.

Furthermore, Jaber [15] suggested a video steganography technique that replaces the traditional LSB algorithm with bitwise AND and OR operations. The private information is encrypted using a key generated by a GA, and the encrypted data is embedded in a cover video. The method boasts a high PSNR range of 60 to 63 dBs. However, it should be noted that the authors have only improved the encryption aspect of the technique and did not present any advancements in the embedding method. Additionally, the experimental results were limited to only two videos. Finally, Job et al. [16] presented a secure data transfer technology based on improved steganography. The method combines the use of Elliptic Curve Cryptography (ECC) for data encryption with improved steganography techniques utilizing a sudoku matrix and GA for data embedding into cover videos. The process involves selecting a frame of the video at a time, computing the PSNR, and reinserting the frame if the PSNR is high. The results of the study show that when a short message such as "BYE, WE, US" is hidden in a frame, the PSNR is between 84 and 88 dBs. However, the study does not present the outcome of using a longer message or file as the secret data, and the results are based on testing with only one frame.

## 3  Advanced Encryption Standard

The Advanced Encryption Standard (AES) is a widely-used symmetric encryption algorithm that was first adopted by the US government in 2002. It was designed to replace the older Data Encryption Standard (DES) algorithm, which had become increasingly vulnerable to attack due to its relatively small key size. AES uses a block cipher, which means that it encrypts data in fixed-size blocks, typically 128 bits in length. It also uses a key, which is a string of bits that is used to encrypt and decrypt the data. The length of the key can vary, with options ranging from 128 bits to 256 bits [17]. The encryption process in AES involves a series of rounds, in which the data is transformed using a series of mathematical operations. The number of rounds that are performed depends on the key size, with longer keys requiring more rounds. This makes AES more secure, as it requires more computational resources to crack.

Several different modes of operation can be used with AES, including Electronic Codebook (ECB), Cipher Block Chaining (CBC), Cipher Feedback (CFB), and Output Feedback (OFB). Each of these modes has its unique characteristics and is suitable for different types of applications. In terms of security, AES is considered to be very strong. It has resisted all known attacks, including brute-force attacks, which try all possible key combinations to decrypt the data. However, like all encryption algorithms, it is not completely foolproof and future attacks may be developed that could compromise its security. Overall, AES is a reliable and secure encryption algorithm that is widely used in a variety of applications, including internet communication, file storage, and secure data transmission. Its strong security and wide adoption make it an important tool for protecting sensitive data.

Here is a simplified block diagram of the Advanced Encryption Standard (AES) algorithm:

- The plaintext data is divided into blocks of fixed size, typically 128 bits.
- The key is used to initialize a key schedule, which generates a series of round keys that will be used in the encryption process.
- The plaintext block is XORed with the first-round key to produce the intermediate state.
- The intermediate state is passed through a series of rounds, each of which consists of four main steps:
    - The state is transformed using a non-linear substitution step, in which each byte of the state is replaced with a predetermined value that is determined by the key.
    - The state is shifted using a row shift step, in which the rows of the state are cyclically shifted to the left.
    - The state is mixed using a column mix step, in which the columns of the state are transformed using a series of mathematical operations.
    - The state is XORed with the round key for that round.

- After the final round, the intermediate state is transformed using a final round function, which consists of the substitution and shift steps, but not the mixing step.
- The final state is XORed with the final round key to produce the ciphertext.
- The ciphertext is output and can be transmitted or stored securely.

To decrypt the ciphertext, the process is reversed, using the same key and key schedule. The ciphertext is XORed with the final round key and transformed using the inverse of the final round function, followed by a series of inverse rounds. The intermediate state is then XORed with the first-round key to produce the plaintext [18,19].

## 4  The Proposed Video Steganography Method

In this research, a novel approach to video steganography is proposed, utilizing a combination of GA and AES encryption techniques. The proposed method utilizes YUV videos as carriers and the Y component to embed secret data, to hide QR codes within the video. The method is unique in its reversibility, allowing for the retrieval of hidden data without causing any damage to the original video. The system employs a multi-step process to generate the stego video, including the utilization of GA and AES encryption. The proposed method is a significant contribution to the field of video steganography, as it addresses the challenge of securely sending data over the internet. With its high embedding capacity and efficiency, the method has the potential to be useful in applications that require secure data transmissions, such as military and intelligence operations, finance, and healthcare. In summary, this paper proposes a new and effective video steganography method that can be used to securely transmit data over the internet while maintaining video quality and efficiency.

### *4.1 Carrier Video Encoding Using GA*

The proposed system utilizes the YUV color format for video as a carrier for embedding secret data. This format separates the pixel brightness, represented by the Y value, from the pixel color, represented by the U and V values. To utilize this feature, the video is divided into frames, and the Y values of each frame are organized into a matrix of dimensions (height X width). This matrix is then processed through a GA for data embedding. This approach leverages the properties of the YUV color format to efficiently conceal data while minimizing the impact on visual quality. The encoding process of the proposed video steganography method is shown in Fig. 1.

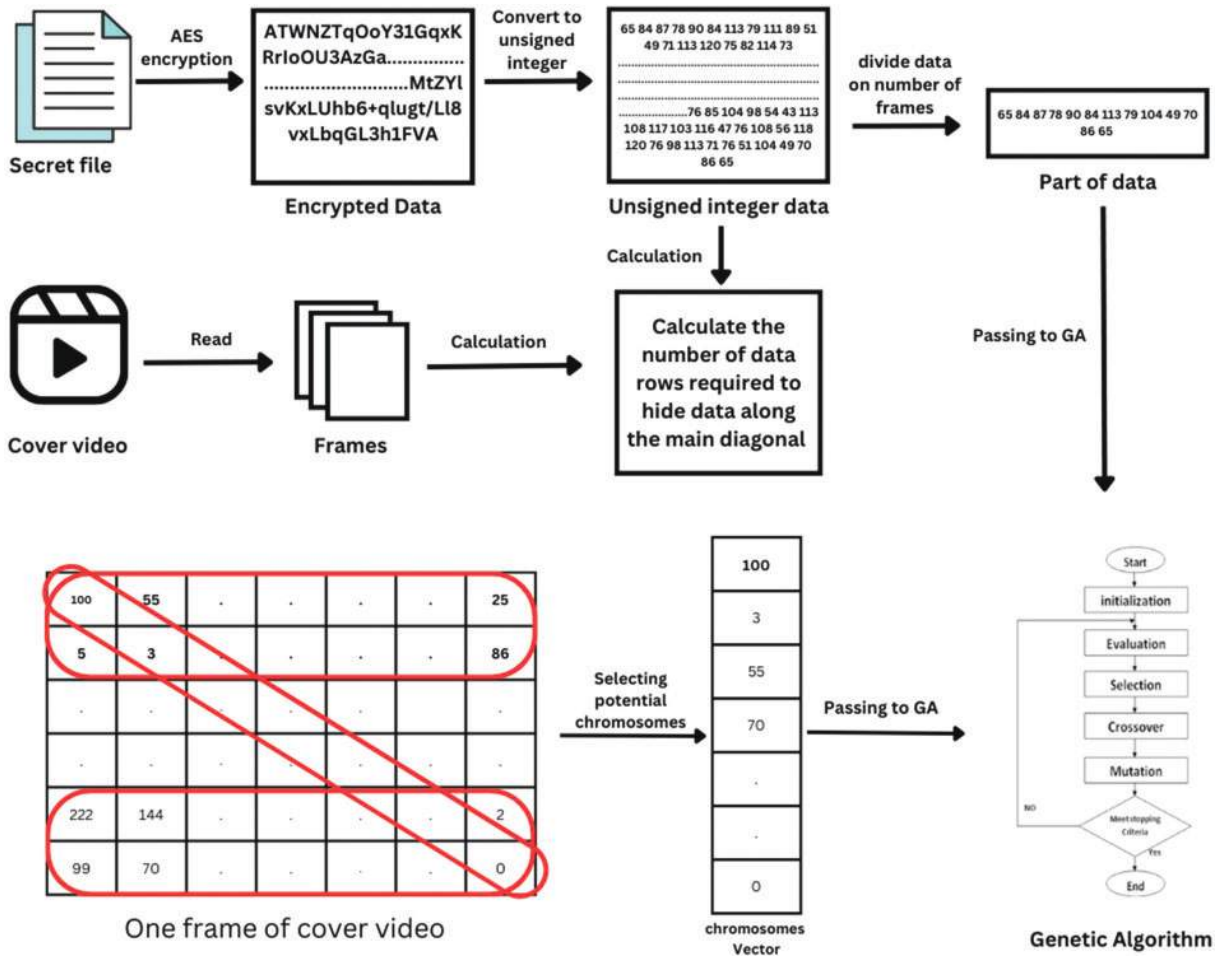

**Figure 1:** The encoding process of the proposed video steganography method

To begin, the confidential data is encrypted with AES using a 128-bit key and is subsequently converted to unsigned integers. This data is then divided and embedded into each frame of the cover video. A calculation method is used to determine the number of pixels needed for data embedding. The selected rows are calculated based on the size of the confidential data.

$$R = \frac{\left(\dfrac{L_M}{F}\right)}{W * 0.1} \tag{1}$$

$L_M$ = Length of embedded data, F = Number of frames, and W = Width of the frame.

The equation determines the number of rows required for embedding data along the main diagonal. If the secret data's length divided by the number of frames is less than 10% of the frame width, only the main diagonal is used. Otherwise, two extra width rows are taken, and if still insufficient, the second and last rows −1. To maintain undetectability, only 10% of each row is used. The system selects the necessary number of rows based on the data size, taken equally from the top and bottom of each frame.

Next, array of potential selected pixels is created and for each pixel one chromosome is created and passed to GA. Fig. 2 illustrate the proposed GA.

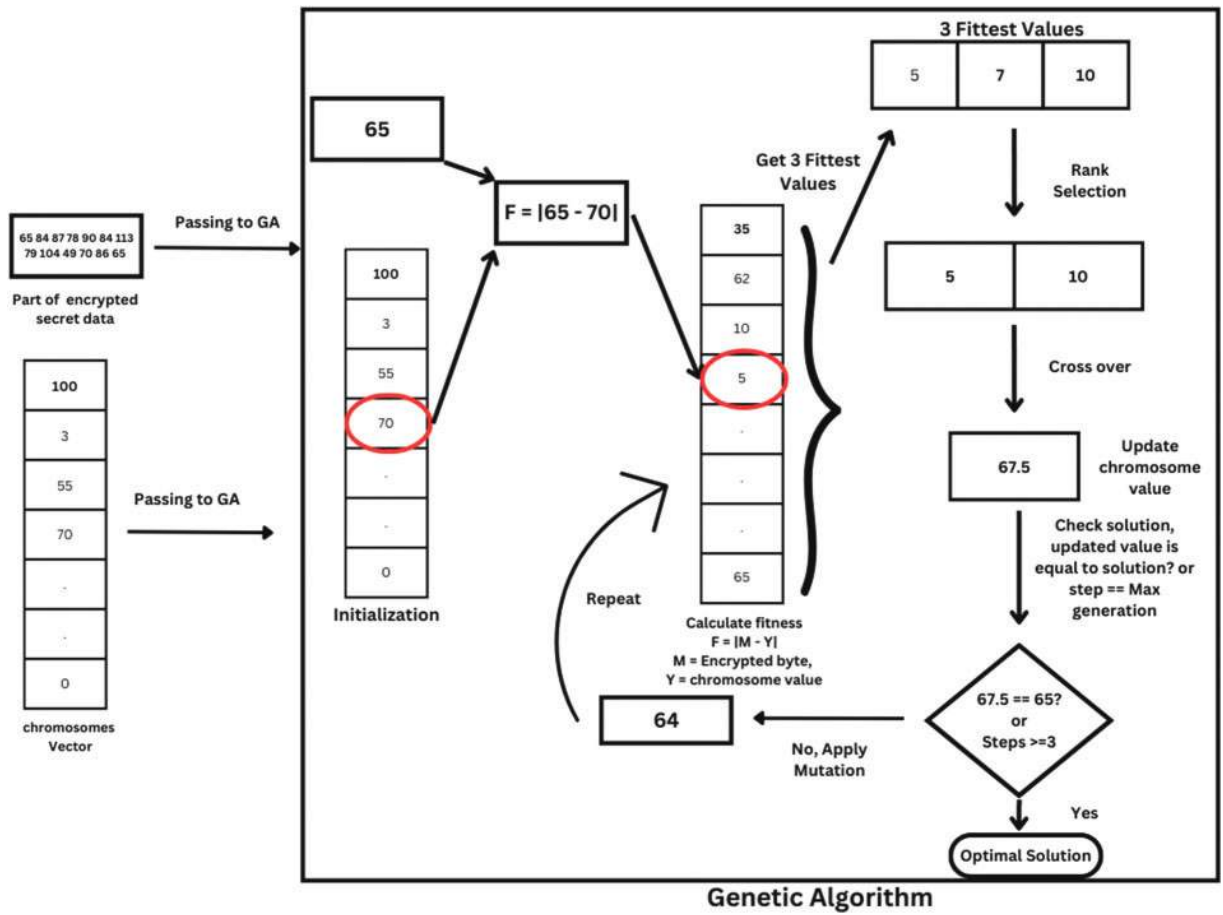

**Figure 2:** The diagram of the proposed GA for video steganography

The first step of GA involves calculating chromosome fitness and selecting the initial population using rank selection, a type of parent selection strategy in GA. The fitness is determined as follows:

$$Fitness = \sum_{i=0}^{L_Y} |M - Y(i)| \tag{2}$$

M = Encrypted character of embedded data, Y = Chromosome Y value, and $L_y$ = Length of chromosomes.

The next step is evolving the population through selection, crossover, and mutation to converge towards the solution by creating offspring from the best chromosomes in each generation.

- Selection: The top three fittest chromosomes were identified, and two were randomly selected as parents for crossover.
- Crossover: Then the two values will be added or subtracted from the result of multiplying the crossover rate (0.5) by the subtraction of the desired solution from the selected value. The crossover is determined as follow:

  $$R = R \pm 0.5 * (DS - R) \tag{3}$$

  where R = chromosomes value, DS = Desired Solution (encrypted secret data value).

- Mutation: The process checks whether the optimal solution has been reached or if the maximum generation has been exceeded. If not, the mutation operator is applied to the chromosome. A mutation rate of 0.05 is used, where the chromosome value is multiplied by the mutation rate, and the resulting value is randomly added or subtracted from the original value of the chromosome to introduce diversity. The Mutation is determined as follow:

  $$R = R \pm 0.05 * R \tag{4}$$

  where R = chromosome value.

The described process will be applied iteratively for each encrypted secret data byte, one at a time.

The program will utilize an inverse method of reading to write the current embedded frame to a new file. This method utilizes mathematical equations to ensure the data is written to its appropriate location. As only the Y value is updated, the U and V values will remain unaltered, and the three values will be combined to form a new pixel.

The utilization of a GA in a data-hiding method may result in varying outcomes due to its randomized nature. To mitigate this, it is recommended to store the modified data in a file to recover the secret data. The information regarding the selected pixels, including the frame number, x, and y coordinates, along with the difference between the original Y values and the modified values, is documented and saved in a CSV file at the end of each processing frame. This file, along with the stego-encoded video file, can then be transmitted to the recipient. For enhanced security measures, the CSV file can be encrypted before its delivery. The flowchart of the proposed video steganography method is depicted in Fig. 3.

Exploration and exploitation are important in search algorithms like genetic algorithms. Exploration is searching the solution space for potential solutions while exploitation uses the best solution to refine the search. In genetic algorithms, initialization, crossover, and mutation represent exploration while selection and evaluation represent exploitation. The balance between exploration and exploitation is crucial for success.

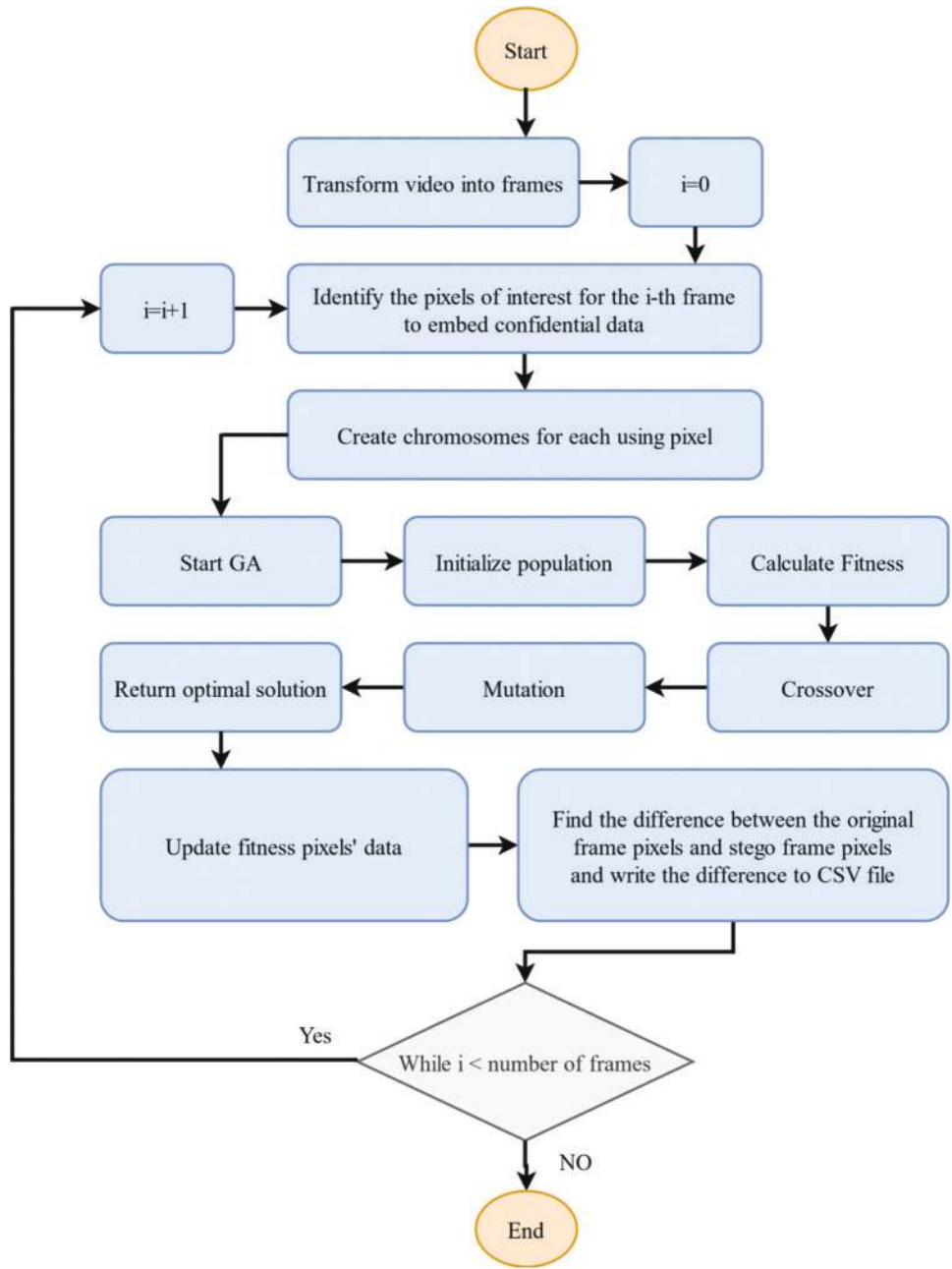

**Figure 3:** The flowchart of the proposed video steganography method

The proposed video steganography method employs GA as the optimization technique due to its proficiency in effectively managing complex, non-linear problems. Furthermore, GA is adept at navigating high-dimensional search spaces with multiple objectives, making it well-suited for our task of embedding data into a video while preserving video quality. In addition, GA has been extensively researched and utilized in the field of steganography, demonstrating superior performance compared to other optimization algorithms in various contexts. Despite the availability of many evolutionary algorithms, the decision was made to utilize GA due to its wide prevalence and demonstrated success in similar applications.

### 4.2 Decoding Stego Video

To retrieve confidential information from a carrier video, the receiver must possess three crucial components: a stego video, a password, and a CSV file. The proposed method involves dividing the stego video into individual frames and decoding each frame separately. Utilizing the information contained within the CSV file, the algorithm will identify and extract the pixels containing the secret data, leading to a reduction in processing time compared to traditional encoding techniques that examine the entire image.

Upon extracting the secret values from the stego video, the extracted values and the corresponding information stored in the CSV file will be subjected to one of the following Eqs. (5) and (6).

If CSV value greater is positive, then:

$$D = CSV + Y \tag{5}$$

Otherwise

$$D = Y + CSV \tag{6}$$

CSV means the stored file in CSV file while Y indicates the extracted Y value from the stego video.

The outcome of the aforementioned mathematical equations and the secret password are fed into the AES encryption scheme, which performs decryption and retrieves the original file. The recovered file is identical to the input file that was encrypted during the encoding process. Fig. 4 illustrates the decoding process of the proposed video steganography method.

## 5 Performance Evaluation

The proposed method was subjected to empirical evaluation using five YUV videos and three QR images of varying dimensions. The work was written in Java and evaluated on a computer using the following specifications: CPU AMD Ryzen 7 3700x @ 3.59 GHz, encoding time ranged from 3 to 152 s and decoding time ranged from 1 to 2 s.

The assessment of the performance of stego videos was based on two fundamental criteria, namely imperceptibility and embedding capacity. To gauge the quality of the embedded data and determine the disparity between the carrier and stego videos, PSNR was employed as the primary metric. PSNR is a widely adopted measure in the field and has been used in numerous studies [20,21]. The calculation of PSNR is described by Eq. (7).

$$PSNR = 10 * \log_{10} \frac{L^2}{MSE} \tag{7}$$

$L$ is the number of maximum possible intensity levels, and the $MSE$ is calculated as follows:

$$MSE = \frac{1}{H * W} \sum (P_{(i,j)} - S_{(i,j)}) \tag{8}$$

where $H$ = Height, $W$ = Width, $P$ = Original video frame, and $S$ = Stego video frame

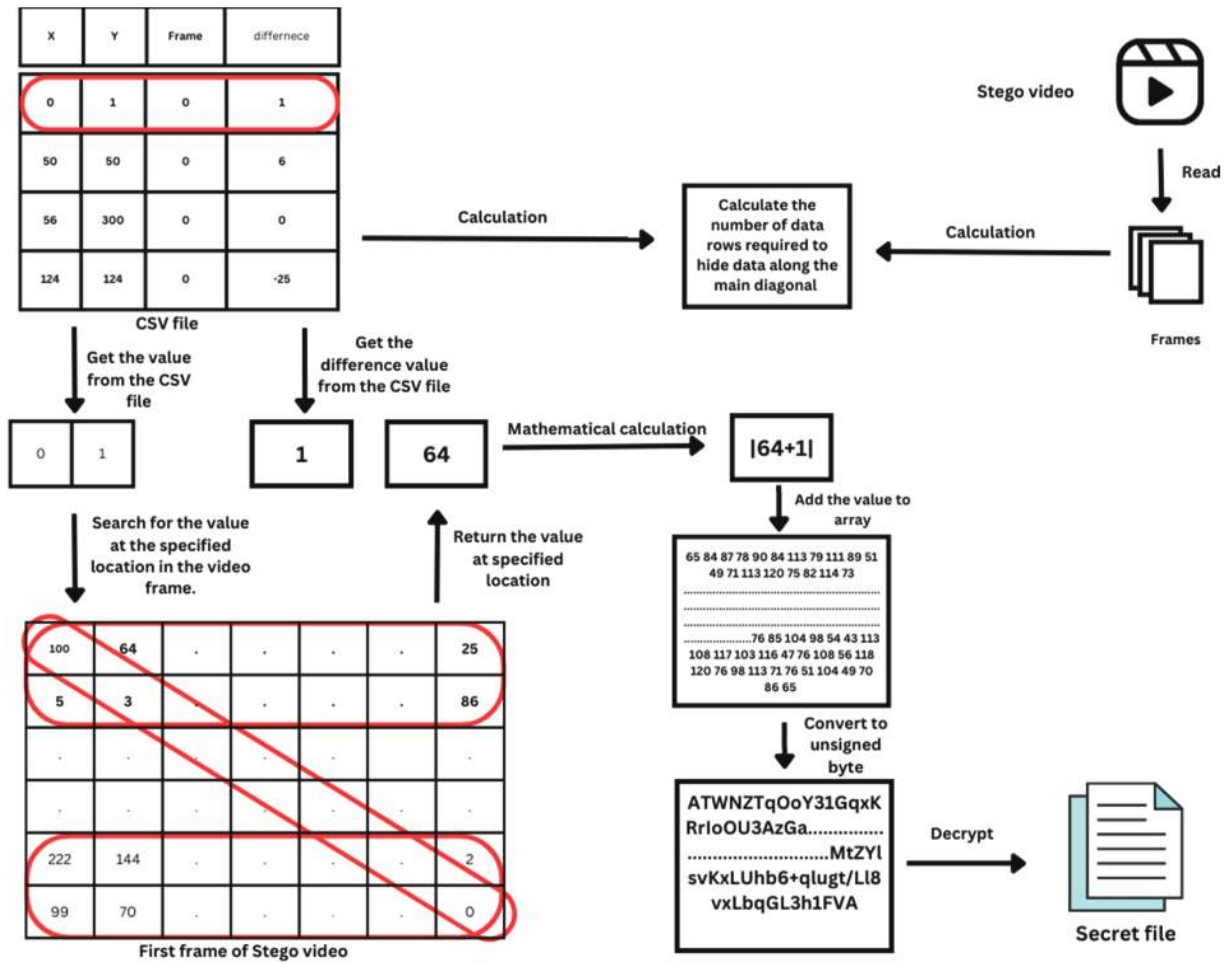

**Figure 4:** The decoding process of the proposed video steganography method

Experimental results are presented in Table 1 to demonstrate the performance of the proposed method. The payload size of the proposed method is determined by the number and size of the frames in the video. This approach ensures that the maximum amount of data can be hidden within the video file without compromising the quality of the video. The results of the proposed method show a PSNR range between 64 and 75, which demonstrates the effectiveness of the approach in preserving the quality of the cover video. Moreover, the encoding and decoding times are efficient, making the proposed method practical for use in real-world applications. Overall, this new video steganography method with a GA provides an innovative solution for secure data hiding within video files with minimal impact on video quality and processing time.

**Table 1:** Experimental results of the proposed video steganography method

| Cover video | Size | Frames | Secret image | Encoding time/seconds | PSNR | MSE |
|---|---|---|---|---|---|---|
| 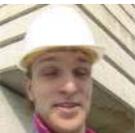 Foreman | $352 \times 288$ | 300 | 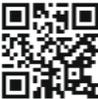 $63 \times 63$ (15.5 KB) | 3 | 75.4 | 0.002 |
| 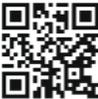 Foreman | $352 \times 288$ | 300 | 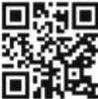 $138 \times 138$ (74.4 KB) | 19 | 67.3 | 0.014 |
| 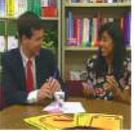 Foreman | $352 \times 288$ | 300 | 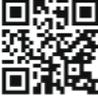 $188 \times 188$ (138 KB) | 1:46 | 64.6 | 0.027 |
| 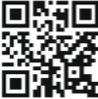 Paris | $352 \times 288$ | 1065 | $63 \times 63$ (15.5 KB) | 7 | 75.9 | 0.001 |
| 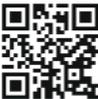 Paris | $352 \times 288$ | 1065 | $138 \times 138$ (74.4 KB) | 17 | 72.5 | 0.003 |
| 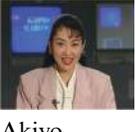 Paris | $352 \times 288$ | 1065 | $188 \times 188$ (138 KB) | 32 | 71 | 0.005 |
| 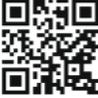 Akiyo | $352 \times 288$ | 300 | $63 \times 63$ (15.5 KB) | 4 | 73 | 0.004 |
| 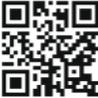 Akiyo | $352 \times 288$ | 300 | $138 \times 138$ (74.4 KB) | 24 | 67 | 0.016 |

(Continued)

**Table 1 (continued)**

| Cover video | Size | Frames | Secret image | Encoding time/seconds | PSNR | MSE |
|---|---|---|---|---|---|---|
| 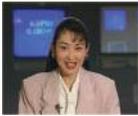 Akiyo | 352 × 288 | 300 | 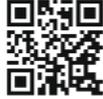 188 × 188 (138 KB) | 1:30 | 64.5 | 0.02 |
| 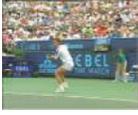 Stefan | 352 × 288 | 90 | 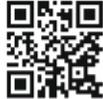 63 × 63 (15.5 KB) | 5 | 69.8 | 0.49 |
| 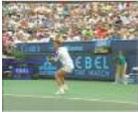 Stefan | 352 × 288 | 90 | 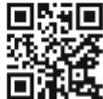 138 × 138 (74.4 KB) | 48 | 67 | 0.015 |
| 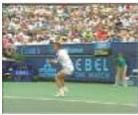 Stefan | 352 × 288 | 90 | 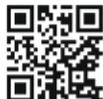 188 × 188 (138 KB) | 2:32 | 63.8 | 0.028 |
| 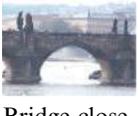 Bridge-close | 352 × 288 | 2001 | 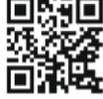 63 × 63 (15.5 KB) | 27 | 70.7 | 0.005 |
| 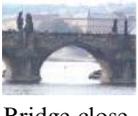 Bridge-close | 352 × 288 | 2001 | 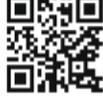 138 × 138 (74.4 KB) | 22 | 72.9 | 0.003 |
| 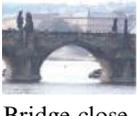 Bridge-close | 352 × 288 | 2001 | 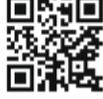 188 × 188 (138 KB) | 32 | 70.7 | 0.005 |

Fig. 5 illustrates the PSNR of the Foreman stego video, which indicates the level of distortion in the stego video as compared to the original video. This measurement serves as a valuable indicator of the quality and fidelity of the stego video, as a higher PSNR value suggests a closer resemblance to the original video in terms of quality. By examining the PSNR value of the Foreman stego video, one can gain insight into the effectiveness of the steganographic method used and its suitability for a particular application.

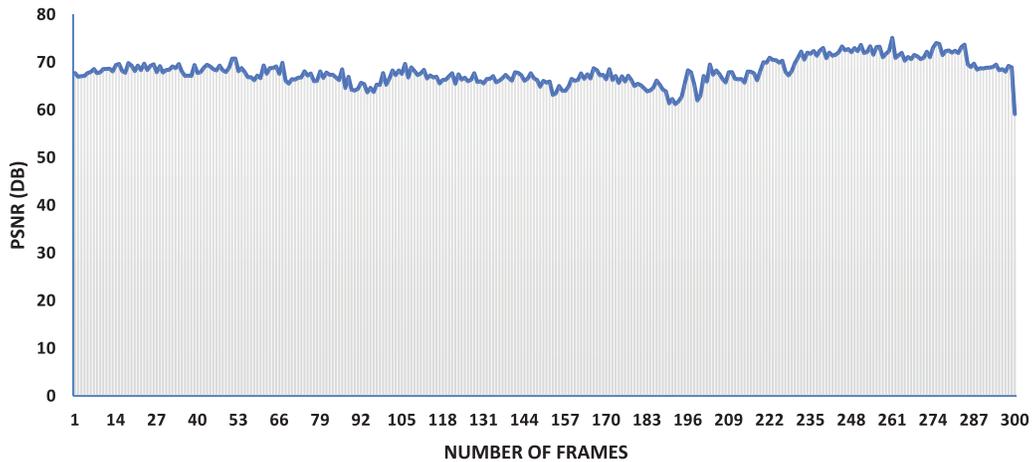

**Figure 5:** The PSNR values of the Foreman stego video using the proposed method

Histograms are graphical representations that show the distribution of colors or intensities in an image. By comparing the histograms of the original and stego frames, it becomes possible to assess the degree of distortion or alteration introduced by the proposed steganographic method used to embed the QR image. In Fig. 6, it is difficult for us to observe a comparison between the histograms of two different frames from a Foreman video. The first histogram represents a randomly selected frame from the original video, while the second histogram depicts a frame from the stego video. In the stego video, a QR image with a size of 138 × 138 pixels (74 KB) was inserted to hide information.

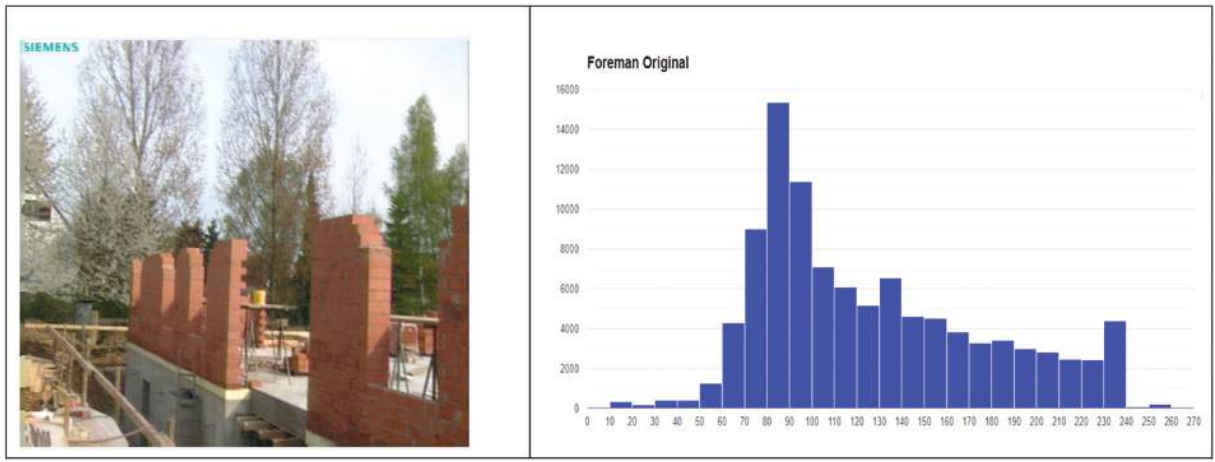

**Figure 6:** (Continued)

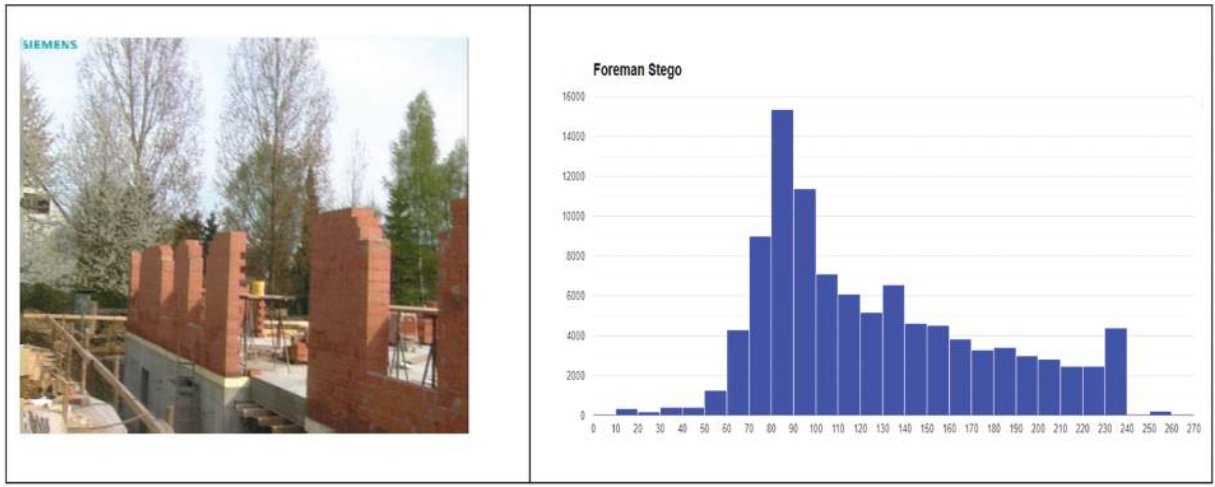

**Figure 6:** The histograms of a single frame from the original Foreman video and its stego counterpart

Additionally, the histogram in Fig. 7 depicts a graphical representation of the distribution of pixel intensities within a frame selected at random from the Stefan video. The first image presents the histogram of the original frame, while the second image presents the histogram of a stego video frame after the insertion of a QR image, which had a size of 138 × 138 pixels and a file size of 74 KB. It is also challenging to discern the differences between the histograms of the two frames. Therefore, it can be inferred that the suggested method is efficient.

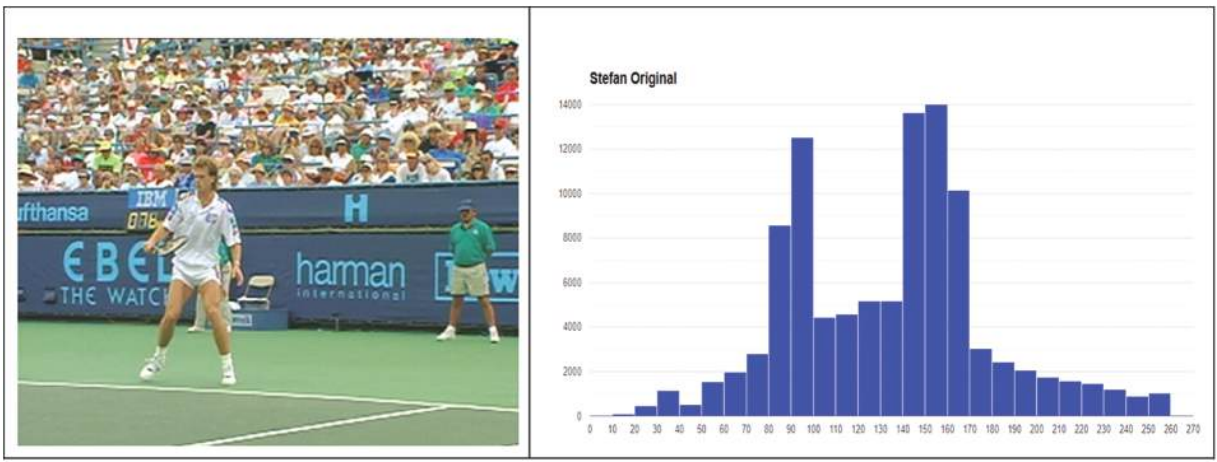

**Figure 7:** (Continued)

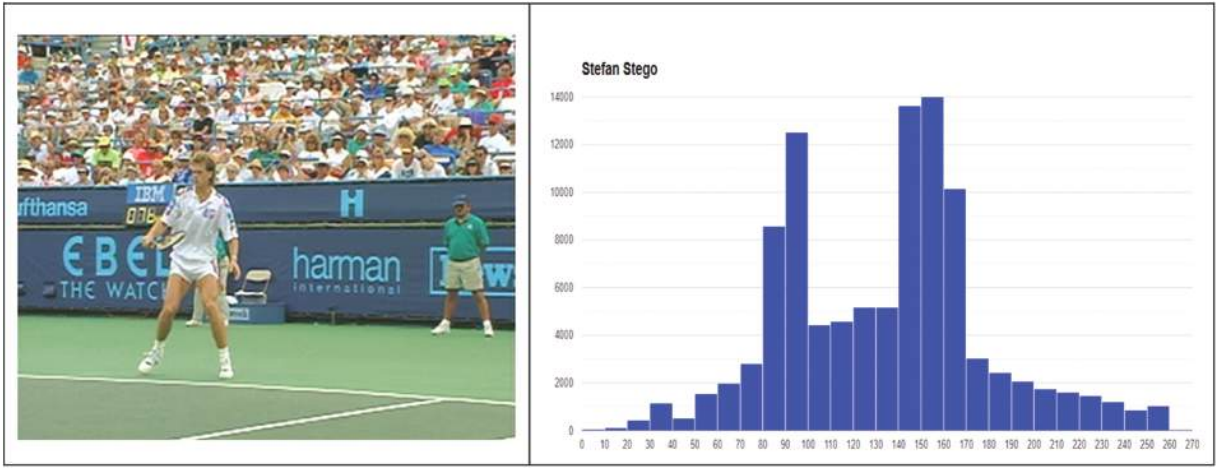

**Figure 7:** The histograms of a single frame from the original Stefan video and its stego counterpart

Table 2 shows that the suggested method has the greatest PSNR rate compared to the methods provided in [10,11] except the Globe video in reference [10]. The proposed method's outcomes in this comparison indicate a PSNR range of 62 to 73, highlighting its effectiveness in maintaining the cover video's quality.

**Table 2:** Comparison of the proposed method with other existing approaches in terms of PSNR

| Cover video | Secret image size (KB) | Video quality (PSNR) dB | | Cover video (AVI) | No. of frames | Video quality (PSNR) dB | |
|---|---|---|---|---|---|---|---|
| | | Proposed method | Ref. [11] | | | Proposed method | Ref. [10] |
| Akiyo | 89 | **64.67** | 63.86 | Tree | 450 | **69.68** | 51.62 |
| Akiyo | 179 | **62.22** | 60.89 | Globe | 107 | 39.57 | **47.74** |
| Foreman | 89 | **66.86** | 63.87 | Computer | 510 | **73.1** | 54.23 |
| Foreman | 179 | **64.31** | 60.87 | | | | |

Table 3 demonstrates that the proposed method has a higher PSNR rate compared to the methods presented in reference [22] when tested on four videos. However, in only Video 1 and Video 3, the PSNR results of reference [7] outperformed ours. Table 4 presents a comparison between our proposed method of video steganography incorporating a GA and a conventional video steganography approach that does not utilize genetic algorithms. Our method leverages the optimization capabilities of the GA to determine the optimal location for data embedding, thus resulting in improved performance.

**Table 3:** Comparative analysis of the proposed method with [7] and [22]

| Videos | Video quality (PSNR) dB | | |
|---|---|---|---|
| | Proposed method | Ref. [7] | Ref. [22] |
| Video 1 | 63.38 | **71.54** | 61.229 |
| Video 2 | **75.50** | 75.14 | 64.931 |
| Video 3 | 70.87 | **72.37** | 64.113 |
| Video 4 | **74.64** | 72.57 | 63.687 |

**Table 4:** Comparative analysis of the proposed method with and without using GA

| Video type | PSNR (Without GA) | PSNR (GA) |
|---|---|---|
| Foreman | 56.4 | **67.3** |
| Paris | 60.7 | **72.5** |
| Stefan | 51.2 | **66.6** |
| Akiyo | 56.5 | **67** |
| Bridge-close | 62 | **72.9** |

## 6 Conclusion

In this paper, a novel video steganography method utilizing GA to identify the optimal location for data embedding within a cover video file was introduced. The GA algorithm was employed to determine the frames within the video file that were most suitable for data embedding, which improved the security of the hidden information and reduced the risk of detection by potential attackers. The proposed approach involved dividing the video data into frames and selecting the Y component for data embedding due to its high sensitivity to changes in color, resulting in a more robust method for data hiding. Additionally, the method optimized the data-hiding capacity of the cover video by utilizing only 10% of its storage space, while maintaining the integrity of the original video. The effectiveness of the approach in preserving the quality of the cover video was evidenced by the PSNR range between 64 and 75, as revealed by the results of the proposed method.

Furthermore, the proposed video steganography method utilizing GA offered substantial benefits in terms of improved imperceptibility and increased embedding capacity. The method was capable of concealing substantial quantities of data without impairing the quality of the carrier video, thus augmenting its security. Nonetheless, as with other evolutionary computation methods, our approach may have required additional time to achieve a high level of accuracy. Nevertheless, we resolved this matter by minimizing the encoding and decoding time to improve the efficiency of the proposed method.

Future research efforts will concentrate on enhancing the hiding capacity of the proposed method by developing a technique for directly embedding data within the stego video file, as opposed to utilizing a separate file for storing the locations of altered pixels. Such an improvement would lead to further enhancements in the security and efficiency of the approach.

**Funding Statement:** The authors received no specific funding for this study.

**Conflicts of Interest:** The authors declare that they have no conflicts of interest to report regarding the present study.